\documentclass[12pt]{iopart}
\usepackage{iopams}

\def\Journal#1#2#3#4{{#4} {\it #1} {\bf #2}, #3 }
\def\a{\alpha}
\def\b{\beta}

\def\e{\epsilon}
\def\s{\sigma}
\def\t{\theta}

\begin{document}

\title{Petrov type I silent universes with G3 isometry group: the uniqueness result recovered}

\author{Lode Wylleman and Norbert Van den Bergh}

\address{Faculty of Applied Sciences TW16, Gent University, Galglaan 2, 9000 Gent, Belgium}

\begin{abstract}
Irrotational dust spacetimes with vanishing magnetic Weyl curvature are called silent universes (Matarrese, Pantano and Saez, 1994). 
%are defined to be irrotational dust spacetimes with vanishing magnetic Weyl curvature. 
The \emph{silent universe conjecture} (Sopuerta 1997, van Elst et al. 1997) states that the only algebraically general silent universes are the orthogonally spatially homogeneous Bianchi I models. In the same paper by Sopuerta this was confirmed for the subcase where the spacetime also admits a group G3 of isometries. However the proof contains a conceptual mistake. We recover the result in a different way. 
\end{abstract}

\pacs{0420Jb,0440Nr}

\section{Introduction}

Irrotational dust $(\omega^a=0, p=0)$ solutions of the Einstein perfect fluid field equations  
\begin{eqnarray}
	R_{ab}-\frac{1}{2}R\,g_{ab}&=&\mu u_au_b
\end{eqnarray}
with vanishing magnetic part $H_{ab}$ of the Weyl tensor \cite{Kramer} were called \emph{silent universes} for the first time by Matarrese, Pantano and Saez \cite{Matarrese1}. The reason for this nomenclatura was that signal exchange in GR can only occur via sound and gravitational waves, none of wich modes are allowed when $p=H_{ab}=0$. From a more mathematical point of view, the evolution equations for the matter density $\mu$, the three non-zero eigenvalues $\theta_1,\theta_2,\theta_3$ of the expansion tensor $\theta_{ab}$ and the two independent eigenvalues (e.g.\ $E_1,E_2$) of the remaining electric part $E_{ab}=C_{acbd}u^cu^d$ of the Weyl tensor, form an autonomous system of  \emph{ordinary} differential equations (see also below); no spatial gradients appear, such that each fluid element evolves indeed as a separate or `silent' universe, once the constraint equations are satisfied by the initial data. This setting looked very appealing towards numerical schemes and simulations in astrophysical and cosmological context, e.g.\ for the description of structure formation in the universe and the study of the gravitational instability mechanism in general relativity, where a clear motivation for taking $p=\omega^a=0,H_{ab}\approx 0$ was given in \cite{Matarrese2}.  

The only allowed Petrov types for silent universes are O, D and I: the Friedman-Robertson-Walker dust metrics exhaust the type 0 class, while all Petrov type D solutions are known explicitly too. They are characterised~\cite{Barnes} by the fact that the Weyl
tensor is degenerate in the same plane as the shear and are given
by the Szekeres~\cite{Szekeres} family, including the subcase of
the Ellis~\cite{Ellis} LRS class II dust models and such well
known examples as the Lemaitre-Tolman-Bondi model and the
orthogonally spatially homogeneous Kantowski-Sachs model. For the algebraically general case (Petrov type I) the situation is rather
different, the only known silent universe being the orthogonally
spatially homogeneous Bianchi I model. 

In two independent papers by Sopuerta \cite{Sopuerta} and van Elst et al. \cite{vanElst}, the propagation of the constraint $H_{ab}=0$ was shown to give rise to a triplet of (in principle) infinite chains of equations, identically satisfied for Petrov type 0, D and spatially homogeneous silent models, but not in general. What happens is that for \emph{non}-spatially homogeneous Petrov type I silent universes the initial values for the $\theta_\a$, $E_\a$ and $\mu$ must lie on a (low-dimensional) algebraic variety, i.e., there exist severe polynomial relations between these variables. As it was conjectured by the authors of \cite{Sopuerta,vanElst}, there are good reasons to believe that the variety is actually empty, i.e., that there exist no non-spatially homogeneous Petrov type I silent models at all, at least for zero cosmological constant. This is the so called `silent universe conjecture', no definitive proof of which has been given until now, mainly because the degree and the number of terms of the polynomials are massive. However, we mention here that in the case of strictly positive cosmological constant, the above variety has been shown to contain three (analogous) two-dimensional components, corresponding with two explicit new families of metrics each \cite{Vandenbergh}. 

In the present paper, we correct a reasoning in \cite{Sopuerta} and reconfirm the silent universe conjecture (for vanishing cosmological constant) in the subcase where the spacetime admits a group G3 of isometries.\\

For silent universes, one can deduce in general (Proposition 3 of \cite{Sopuerta}) that there always exists an orthonormal basis $\{e_0,e_\a\}$ (with $e_0\equiv u$) of common eigenvector fields of the expansion tensor $\theta_{ab}$ and the remaining electric part $E_{ab}=C_{acbd}u^cu^d$ of the Weyl tensor, which are all parallelly transported along the matter flow lines and are hypersurface orthogonal, such that a local coordinate system $(t,x^\b)$ exists for which $e_0%\equiv\partial_0
=\frac{\partial}{\partial t}$ and $e_\a%\equiv\partial_\a
\propto\frac{\partial}{\partial x^\a}$. Below we will use the notation $\partial_\a$ for $e_\a$ whenever the frame vector fields act as differential operators.

In \cite{Sopuerta} algebraically general silent universes were investigated in such  coordinates, and in section 5  restrictions on the coordinate components of a generic Killing vector field $K$ were obtained. However, in the discussion of Case (iii) a time dependence via $K^z$ in (108) was overlooked and therefore the conclusion that $K^t$ is constant and $K^{x^\a}$ only depends on $x^\a$ does not necessarily hold. But even if this conclusion were valid, a coordinate transformation $x^\a\rightarrow x^{\a\prime}$ which transforms the components of a \emph{fixed} Killing field $K_0$ into constant functions won't necessarily do the same for another Killing field $K$, such that coordinates w.r.t.\ which \emph{every} Killing field has constant coefficients (which was the content of Theorem 1 in \cite{Sopuerta}) do not necessarily exist. 
Herewith the reasoning  leading to the uniqueness of G3 Petrov type I silent universes (Corollary 1 of  \cite{Sopuerta}) breaks down.
    
Fortunately, there is a related but essentially different way to prove the G3 uniqueness  result. The key point is that, contrary to gauge dependent quantities such as the metric components, \emph{invariantly-defined geometric objects} must be invariant under the isometries \footnote[1]{see \cite{Kramer}, $\S\,8.4$  for an account}. For example, scalar invariants such as the matter density and the eigenvalues of Weyl and expansion tensor must by annihilated by any Killing field, and this yields sufficient extra information to prove the key Lemma 3 below. In particular for Petrov type I, we note that the eigenvector fields $e_\a$ of the Weyl tensor and hence the corresponding rotation coefficients $\Gamma_{abc}:=e_a\cdot\nabla_c e_b$ are examples of invariantly-defined geometric objects and hence are preserved under local isometries.

\section{Mathematical setting}

In what follows, $U$ will always denote an open subset of a spacetime for which the conditions in the definition of a silent universe are satisfied everywhere. For such a subset, we will say the spacetime is BI if it is an orthogonally spatially homogeneous Bianchi I model on $U$. The set of smooth functions on $U$ will be denoted ${\cal F}(U)$, and all reasonings will only involve a finite number of polynomial combinations $F_i$ (including 0) of such functions. Because of this, we can always assume $U$ small enough such that for all couples $(F_i,F_j)$
either $F_i(p)=F_j(p), \forall p \in U$ (further denoted by $F_i=F_j$) or $F_i(p)\neq F_j(p),\forall p \in U$ (denoted by $F_i\neq F_j$).

For silent universes, the only remaining dynamical variables are the matter density $\mu$, the expansion eigenvalues $\t_\a$, the Weyl eigenvalues $E_\a$, the purely spatial Ricci~rotation coefficients $2q_\a\equiv-e_\a\cdot\nabla_{\a-1}e_{\a-1}$ and $2r_\a\equiv e_\a\cdot \nabla_{\a+1}e_{\a+1}$, and $m_\a:=\partial_\a(\mu)$. In these variables, the first contracted Bianchi identity reads $\partial_0\mu=-\mu\,\theta$ (with $\t$ the expansion scalar), and for the derivatives of the expansion and Weyl eigenvalues $E_\a$, resp.\ $\t_\a$ one has \cite{Vandenbergh}:
\begin{eqnarray}
\label{d0H}\partial_0 \t_\a & = &\,- E_\a-\t_\a^2-\frac{\mu}{6}, \\
\label{d0E}\partial_0 E_\a &=&\,
2\t_{\a+1}\,x_{\a-1} - 2\t_{\a-1}\,x_{\a+1}-\frac{\mu}{2}\,\s_\a,\\
x_\a\label{dHdiag}\partial_\a \t_\a&=&\, 6 E_\a(-h_{\a+1}\,q_\a+ h_{\a-1}\,r_\a)
-\frac{h_\a}{2}m_\a, \\
\label{dHoff}\partial_\a \theta_{\a+1} & = &\,-4 h_{\a-1}\, r_\a,\quad 
\partial_\a \t_{\a-1}  = -4 h_{\a+1}\, q_\a,\\
\label{dEoff}\partial_\a E_{\a+1} &=&\,  -4 x_{\a-1}\,r_\a -\frac{m_\a}{6},\quad 
\partial_\a E_{\a-1}  = -4 x_{\a+1}q_\a -\frac{m_\a}{6}.
\end{eqnarray} 
Here the expressions have to be read modulo 3 ($\a$ running from 1 to 3), $\s_\a~\equiv~ \t_\a-1/3\,\t$ denote the shear eigenvalues and $2h_\a:=~\t_{\a+1}~-~\t_{\a-1}$, $2x_\a:=~E_{\a+1}~-~E_{\a-1}$. 
We remark that a particular result stated in these variables actually yields a triple of results, by cyclic permutation of the indices. We will use this in Lemma 1(a) and Lemma 3.  
For most of our purposes, it will be favorable to work with the following linear combinations of $\t_\a$ and $E_\a$: 
\begin{equation}
  x_1,\quad h_1,\quad \e_1:=-\frac{3}{2}E_1,\quad s_1:=-\frac{3}{2}\s_1,\quad b_1:=\frac{\t_2+\t_3}{2}.
\label{EHs}	
\end{equation}
The expressions for the derivatives of these variables (in terms of themselves and $q_\a,r_\a,m_\a$) can be readily deduced from (\ref{d0H})-(\ref{dEoff}). 
We shall frequently make use of the following three results:
\begin{enumerate}
\item[(r1)] $\theta_\a=\theta_\b$ ($h_1=0,s_1=h_1,s_1=-h_1$) implies $E_\a=E_\b$ ($x_1=0,\e_1=x_1,\e_1=-x_1$), 
see \cite{Barnes};
\item[(r2)] if all $q_\a$ and $r_\a$ vanish then the spacetime is BI, see \cite{vanElst,Mars}; 
\item[(r3)] the only equilibrium point of the autonomous dynamical system formed by  $\partial_0\mu=-\mu\,\theta$,(\ref{d0H}) and (\ref{d0E}) is $(E_\a~=~0,\theta_\a~=~0)$, see  \cite{Sopuerta,Bruni}.
\end{enumerate}

\section{Recovering the uniqueness result for G3 silent universes}

{\bf Lemma 1 (special cases).} If (a) $\t_\a=0$ ($b_1=s_1,b_1=-h_1,b_1=h_1$),
(b) $b_1=0$ or (c) $x_1s_1=\e_1h_1$ on $U$, then the Petrov type is 0 or D.\\ 
{\bf Proof.} (a) It suffices to prove this for $\a=1$ (cf.\ supra). Consecutive time derivatives of $b_1=s_1$, substituting previously obtained equations in each step, yield $\mu=4\e_1$, $-\e_1s_1+x_1h_1=0$ and $(\e_1-x_1)(\e_1+x_1)=0$, i.e.\ $E_1=E_3$ or $E_1=E_2$. (b) The same procedure for $b_1=0$ yields  $\mu = -6h_1^2-2\e_1$ and $x_1h_1=0$, hence $E_2=E_3$ by (r1). (c) For $x_1s_1=\e_1h_1$ this procedure immediately yields $x_1(s_1-h_1)(s_1+h_1)=0$, so again the Petrov type is 0 or D by (r1).$\hfill\opensquare$

{\bf Lemma 2.} An algebraically general silent universe $U$ for which $s_1=c\,h_1$, with $c$ a constant function, is necessarily BI.

{\bf Proof.} The same procedure as in Lemma 1, applied on $s_1=c\,h_1$, yields $\e_1=(c^2-1)h_1^2+c\,x_1$, $h_1(c-1)(c+1)(c\,h_1^2-b_1h_1+x_1)=0$. By (r1), the last equation implies $x_1=h_1(b_1-c\,h_1)$, and hence $\e_1=h_1(c\,b_1-h_1)$ by substition. Two further time evolutions yield $\mu=-h_1^2-2ch_1b_1+3b_1^2$ and an identity. Substituting the obtained expressions for $s_1,\e_1,x_1,\mu$ into (\ref{dHdiag})-(\ref{dEoff}) yields 3 homogeneous linear systems of 5 equations in the variables $\partial_\a b_1,\partial_\a h_1,q_\a,r_\a,m_\a$ (for each $\a$ separately), the coefficients of which depend polynomially on $b_1$ and $h_1$. The product of the 3 corresponding determinants is constantly proportional to $h_1^{12}(c-1)^8(c+1)^8(h_1-b_1)^2(b_1+h_1)^2$, and this is non-zero by (r1) and Lemma 1(a). Hence the linear systems only allow the trivial solution; the result follows by (r2).$\hfill\opensquare$   

{\bf Lemma 3.} When a vectorfield of the form $\partial_\a+g_\a \partial_0$, defined on an algebraically general silent universe $U$ ($g_\a\in {\cal F}(U)$) annihilates the scalars $\mu$ and (\ref{EHs}), then $g_\a= 0$.

{\bf Outline of proof.} It suffices to prove this for $\a=1$ (cf.\ supra), and we set $X:=\partial_1+g_1\partial_0$. 
By substituting the annihilation condition $m_1 = g_1\,\mu\,(3b_1-s_1)$ for $\mu$ into  $X(x_1)=X(h_1)= X(\e_1)=X(x_1)=x_1X(s_1)=X(b_1)= 0$ respectively, one gets five equations $eq_1,\ldots,eq_5$ which form a homogeneous linear system in the variables $q_1,r_1,g_1$, the coefficients of which depend polynomially on $\mu,x_1,h_1,\e_1,s_1,b_1$. 

Assume $g_1\neq 0$. This implies $\textrm{det}_{1}=\textrm{det}_{2}=\textrm{det}_{3}=0$, where $\textrm{det}_i$ is the determimant corresponding to the linear system $(eq_1,eq_2,eq_i)$, $i=3,4,5$. The combinations $\textrm{det}_{2}-\frac{1}{3}h_1\textrm{det}_1$ and $6\e_1\textrm{det}_{3}-2s_1\textrm{det}_1$ can be factorized as $(x_1s_1-\e_1h_1)p_1$, $(x_1s_1-\e_1h_1)p_2$ respectively, such that $p_1=p_2=0$ by Lemma 1(c). To complete the proof we will use three additional relations $p_3=p_4=p_5=0$, where $p_3$ is defined by $\partial_0p_1+(6b_1-2s_1)p_1:=4\,h_1\,p_3$ and $p_4,p_5$ are the respective polynomials $\partial_0\,p_2+(5b_1-s_1)p_2$ and $\partial_0\,p_3$.   
%($\partial_0(\textrm{det}1)\equiv0$ leads to redundant information in a later stage). 

We scale $\mu,\e_1,x_1$ with $h_1^2$ and $b_1,s_1$ with $h_1$; this comes down to specializing  $h_1=1$ in $p_1,\ldots,p_5$, and we will denote the resulting variables and polynomials with a bar. The respective resultants w.r.t.\ $\overline{\mu}$ of the couples $(\overline{p}_1,q)$ with $q=\overline{p}_2,\overline{p}_3,\overline{p}_4,\overline{p}_5$ yield four irreducible polynomial relations $u_1= u_2= u_3=u_4=0$  for the variables $\overline{\e}_1,\overline{x}_1,\overline{b}_1,\overline{s}_1$. 
Next, the resultants w.r.t.\ $\overline{e}_1$ of the couples $(u_3,u_i)$, $i=1,2,4$, are respectively of the form $\overline{x}_1\,\overline{b}_1\,v_1(\overline{x}_1,\overline{s}_1,\overline{b}_1)$, $\overline{x}_1\,\overline{b}_1^{2}\,v_2(\overline{x}_1,\overline{s}_1,\overline{b}_1)$ and $\overline{x}_1\,\overline{b}_1^{2}\,v_3(\overline{x}_1,\overline{s}_1,\overline{b}_1)$, such that $v_1=v_2=v_3=0$ by Lemma 1(b). Finally, the resultants w.r.t.\ $\overline{x}_1$ of $(v_1,v_2)$ and $(v_1,v_3)$ yield
\begin{eqnarray}\label{PQ}	
(\overline{b}_1-\overline{s}_1)^2(\overline{b}_1+\overline{s}_1)^2P_1\,P_2=0,&\quad&(\overline{b}_1-\overline{s}_1)^2Q_1\,Q_2=0
\end{eqnarray} 
respectively, where $P_1,P_2,Q_1,Q_2$ and $\overline{b}_1+\overline{s}_1$ are different and irreducible polynomials in $\overline{b}_1$ and $\overline{s}_1$. By Lemma 1(a) the factor $(\overline{b}_1-\overline{s}_1)^2$ in both left hand sides of (\ref{PQ}) can be stripped off.  Since the remaining expressions have no common factors, their resultant w.r.t. $\overline{b}_1$ must yield a non-trivial polynomial relation $P(\overline{s}_1)=0$ \footnote[2]{see e.g.\ \cite{CLO}, pp.\ 159. More explicitly, this can be checked by specializing e.g.\ $s_1=2$ in the remaining expressions: the degrees of $b_1$ don't drop, and calculating the specialized resultant gives a non-zero number, such that the original resultant cannot be identically 0 (see also \cite{Mars} for this argument).}. Hence $\overline{s}_1=\frac{s_1}{h_1}$ is constant on $U$, such that the spacetime is BI by Lemma 2. But then $\partial_1\propto\frac{\partial}{\partial x^1}$ annihilates all $E_\a$, $\theta_a$ and $\mu$, and hence the same is true for $g_1\partial_0=(\partial_1+g_1\partial_0)-\partial_1$, which leads to a contradiction with (r3). We conclude that our assumption $g_1\neq 0$ was false and this finishes the proof.$\hfill\opensquare$\\ 

{\bf Remark.} The purpose of taking the combinations $p_1,p_2,p_3$ instead of  $\textrm{det}_{2},\textrm{det}_{3}$ and $\partial_0 p_1$ from which they are derived, is a lowering of the degree from the start. This results in a reduction of the total computation time by a factor of $10^2-10^3$.\\  

{\bf Theorem.} An algebraically general silent universe $U$ with a group G3 of isometries is necessarily a BI spacetime.

{\bf Proof.} Since the Petrov type is I, the orthonormal basis $\{e_0,e_\a\}$ is uniquely determined (principal Weyl tetrad); for any point $p\in U$, the dimension of the linear isotropy group of $p$ is 0 and hence the dimension of the orbit ${\cal O}_p$ under the supposed G3 isometry group is 3 \footnote[3]{see \cite{Kramer}, $\S\, 8.3$, 8.4 and 9.2 for an account}.  
This implies the existence of three functionally independent Killing vectorfields $K_i$ on $U$, $i=1\ldots3$. Writing the expansions $K_i=K_i^0\,\partial_0+K_i^\a\,\partial_\a=K_i^a\, \partial_a$ w.r.t. the principal Weyl tetrad in block matrix form:
\begin{eqnarray}\label{K}
	K=\left[\begin{array}{c}K_1\\K_2\\K_3\end{array}\right]& =&K^0\partial_0+A\left[\begin{array}{c}\partial_1\\ \partial_2\\ \partial_3\end{array}\right]=M\left[\begin{array}{c}\partial_0\\ \partial_1\\ \partial_2\\ \partial_3\end{array}\right]
\end{eqnarray}
where $K^0\in{\cal F}(U)^{3\times 1}$, $A\in{\cal F}(U)^{3\times 3}$ and $M\in{\cal F}(U)^{3\times 4}$, we thus have rank($M$)= 3 and hence rank($A$)= 2 or 3 everywhere on $U$.
\vspace{.5cm}\\
(i) rank$(A)$=2.\\
In this case, there exist functions $f_1,f_2,f_3\in{\cal F}(U)$ such that $\sum_{i=1}^3 f_i\,A_{i*}=0$, where $A_{i*}$ denotes the $i^{th}$ row of $A$. Applying the same linear combination to (\ref{K}) we get
\begin{equation*}
	\varphi\,\partial_0\equiv(\sum_{i=1}^3 f_i\,K^0_i)\,\partial_0=\sum_{i=1}^3 f_i\,K_i
\end{equation*}
where the coefficient $\varphi$ nowhere vanishes on $U$ because rank$(M)$=3. Hence, since the $K_i$ annihilate all Weyl and expansion eigenvalues, the same is true for $\partial_0$, which leads again to a contradiction with (r3). Hence this case cannot occur.
\vspace{.5cm}\\
(ii) rank$(A)=3$.\\
In this case $A$ is everywhere invertible on $U$. When applying the inverse, say $C$, on the left of (\ref{K}), the $\a$-component of the resulting equations yields
\begin{eqnarray}
	\partial_\a+g_\a \partial_0&=&C_{\a i}K_i 
\end{eqnarray}
with $g_\a:=C_{\a i}K^0_i$. Hence $\partial_\a+g_\a \partial_0$ ($\a=1,2,3$) annihilate all scalar invariants, and this implies $g_\a\equiv 0$ by Lemma 3. Thus all $\partial_\a$ annihilate (in particular) $\theta_{\a+1}$ and $\theta_{\a-1}$, which forces all $r_a$ and $q_\a$ to vanish by (\ref{dHoff}) and (r1); the result follows by (2).$\hfill\opensquare$ 
\vspace{1cm}\\
{\bf Aknowledgement}
\vspace{.5cm}\\ 
LW is a research assistant supported by
the Fund for Scientific Research Flanders (F.W.O.), e-mail:
lwyllema@cage.ugent.be. He wants to thank  R Vera and M Mars for pointing out that the proof of the G3 uniqueness result for silent universes had to be revised. He also wants to thank J. Carminati for his hospitality at Deakin University, Geelong.

\end{document}